# Comment: Monitoring Networked Applications With Incremental Quantile Estimation

**Earl Lawrence, George Michailidis and Vijayan N. Nair**

*Abstract.* Our comments are in two parts. First, we make some observations regarding the methodology in Chambers et al. Second, we briefly describe another interesting network monitoring problem that arises in the context of assessing quality of service, such as loss rates and delay distributions, in packet-switched networks.

## 1. ON THE CHAMBERS ET AL. METHODOLOGY

The Chambers et al. paper proposes an interesting methodology for estimating quantiles in the context of a network application with limited local storage and computing capabilities. Fast and efficient algorithms for estimating quantiles are needed in many commercial and scientific applications. Commercial database systems often use equidepth histograms (which involve computation of quantiles) to select appropriate execution plans for query optimization [4]. Quantiles are also needed for computing association rules in data mining [11] and for quick and easy assessment of data quality [2]. Thus, the methodology developed in the paper can be used in many other applications as well.

*Earl Lawrence is in the Statistical Sciences Group, MS F600, Los Alamos National Laboratory, Los Alamos, New Mexico 87545, USA e-mail: earl@lanl.gov. George Michailidis is Associate Professor, Department of Statistics, University of Michigan, Ann Arbor, Michigan 48109, USA e-mail: gmichail@umich.edu. Vijayan N. Nair is Professor, Department of Statistics, University of Michigan, Ann Arbor, Michigan 48109, USA e-mail: vnn@umich.edu.*

### 1.1 Computational Issues

The basic incremental quantile estimation (IQE) algorithm in the paper, presented in Section 3, converts the data buffer **D** to an empirical CDF $F_D(x)$. The required sorting of the data takes $\mathcal{O}(N \log(N))$ operations and in some instances up to $\mathcal{O}(N^2)$. Further, the operations need to be repeated every time new data are acquired, in order to obtain the new CDF.

There has been considerable work on efficient algorithms for quantile estimation in the computer science (CS) literature that may be useful in improving the above computations. The comparison-based algorithm in [1] finds the quantiles for a given dataset in $\mathcal{O}(N)$ operations; more precise bounds on the number of comparisons needed were reported in [10]. Algorithms for estimating quantiles in the presence of new data records, the problem discussed in this paper, have also been considered in the CS literature. Exact quantiles are expensive to calculate, so algorithms have been devised for finding "$\varepsilon$-approximate" quantiles. An efficient algorithm that maintains a deterministic "sample" of values observed thus far and updates appropriately was presented in [5]. This requires $\mathcal{O}(\log(\varepsilon N)/\varepsilon)$ memory and time. More recent work based on randomized algorithms guarantees $\varepsilon$-approximation with high probability $(1-\delta)$ (see, e.g., [4, 9]).

A different line of research uses ideas from approximation theory (that deals with finding representation for functions) to develop fast algorithms for histogram calculations for streaming data. The concept of *sketches*—simple data structures that allow the reconstruction of the underlying function (e.g., a histogram), plays a key role here. Deterministic as well as randomized algorithms that are fast and efficient have been proposed in the literature for computing histograms (see [3, 6]). It would be interesting to study their usefulness for quantile estimation and compare with the technique proposed here.

### 1.2 Estimation vs. Monitoring

Although the problem in Chambers et al. is motivated as a monitoring problem, the methodology developed focuses primarily on estimation. For monitoring purposes, the statistic should have good performance in detecting changes when they occur. Thus, the procedure should be devised to estimate the current scenario rather than obtain a "good" estimate of the quantile under the baseline (i.e., assuming there has been no change). There is, of course, a huge literature on statistical process control and change-point detection that deals with these issues. We recognize that the authors are familiar with both the issues and the literature. So it would have been appropriate to include some discussion of the relevant issues.

One way to easily adapt the methods proposed in the paper to the monitoring situation is to weight the data from the current period as they become available. This can be done simply through the exponentially weighted moving average (EWMA) statistic

$$Z_t = \lambda X_t + (1 - \lambda) Z_{t-1},$$

where $X_t$ is the estimate of the quantile from the current period.

There are many other alternatives such as the CUSUM statistic. There are also other ways of obtaining the monitoring statistic. For example, one could compute EWMA estimates of the CDF and invert this to obtain the quantiles rather than using the EWMA of the quantiles. Further, there are many interesting questions associated with the design and efficiency of monitoring procedures.

In addition, one needs estimates of baseline variability for monitoring and detecting changes. Thus, we need some discussion of the underlying sources of variation in the baseline environment.

## 2. MONITORING NETWORK QUALITY-OF-SERVICE PARAMETERS USING ACTIVE TOMOGRAPHY

We describe here a different application on network monitoring with which we have been involved and where estimating and monitoring quantiles is of interest. This deals with estimating and monitoring quality-of-service characteristics such as delays and losses in large, packet-switched networks. One of the challenges here is that many service providers do not own the entire network and hence do not have access to all the links. Thus, they cannot access the internal nodes (routers) to collect data on performance measures such as loss rates and delays.

The area of *active network tomography* has emerged as a convenient alternative. This involves "probing" the network from nodes located on its periphery and collecting end-to-end path-level data. The "tomography" problem is to reconstruct the link-level quality-of-service characteristics from the path-level data. The probing is done by transmitting a number of probe packets in a short period of time (on the order of seconds) from source nodes $s$ to destination nodes $d$ (see Figure 1) and recording various end-to-end path-level performance metrics such as losses or delays. Figure 1 shows one-way transmissions. One can also use round-trip data (in which case the sender and receiver are the same nodes). However, the packets may not follow the same path in the two directions; further, the queues and buffers that determine the losses and delays depend on the direction of the traffic, so the analysis of the round-trip data share many of the same characteristics as the one-way traffic in Figure 1.

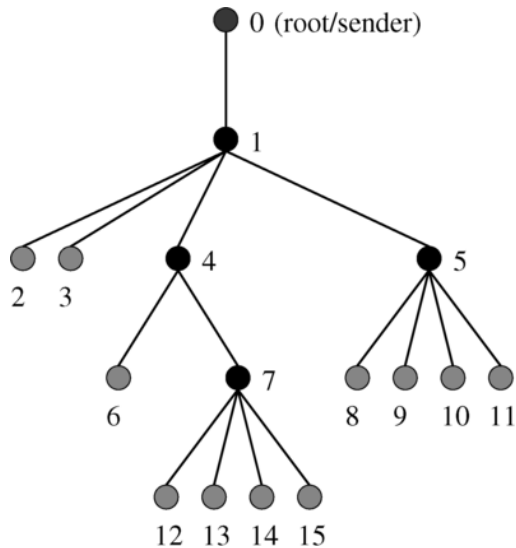

Fig. 1. *A general tree with notation.*

The estimation of link-level characteristics (loss rates and delay distributions) yields interesting cases of large-scale statistical inverse problems. There are many interesting issues in the design of probing experiments, identifiability questions (when can we estimate all of the link parameters?), inference, fast-algorithms and monitoring. See, for example, [8, 12] and references therein. Our goal here is to demonstrate the online monitoring problem that arises through a simulation study using the ns-2 simulator environment [7].

Suppose we are interested in monitoring the performance of the network in Figure 1. This represents the "logical topology" of the subnetwork of interest. We focus on link delay as the performance metric. It is common in the literature to discretize the continuous end-to-end delay data. One then estimates and monitors the delay probabilities associated with the bins (see [8]); that is, the probability of $j$ units of delay on link $k$ is $\alpha_k(j), j = 0, \ldots, b$. These parameters are assumed to be temporally homogeneous. This is a reasonable assumption in this problem as the probing experiments are completed within seconds or minutes. Unlike the application in Chambers et al. with quantiles, the goal here typically involves detecting changes in large delay probabilities.

In our simulation experiment, the capacities of the links are all set to 10 Mbits/sec. Background traffic on each link was generated by forming new TCP and UDP connections with exponential interarrival times and Pareto lifetimes. We probed the network using small constant rate UDP packets (10 per second). The simulation corresponded to one hour of network operations with the background traffic pattern changing on links 4 and 7. In both cases, the changes in traffic (and hence) delay were induced incrementally. The TCP traffic on those links was doubled at 30 minutes and increased by the same amount again at 45 minutes (to triple the original traffic).

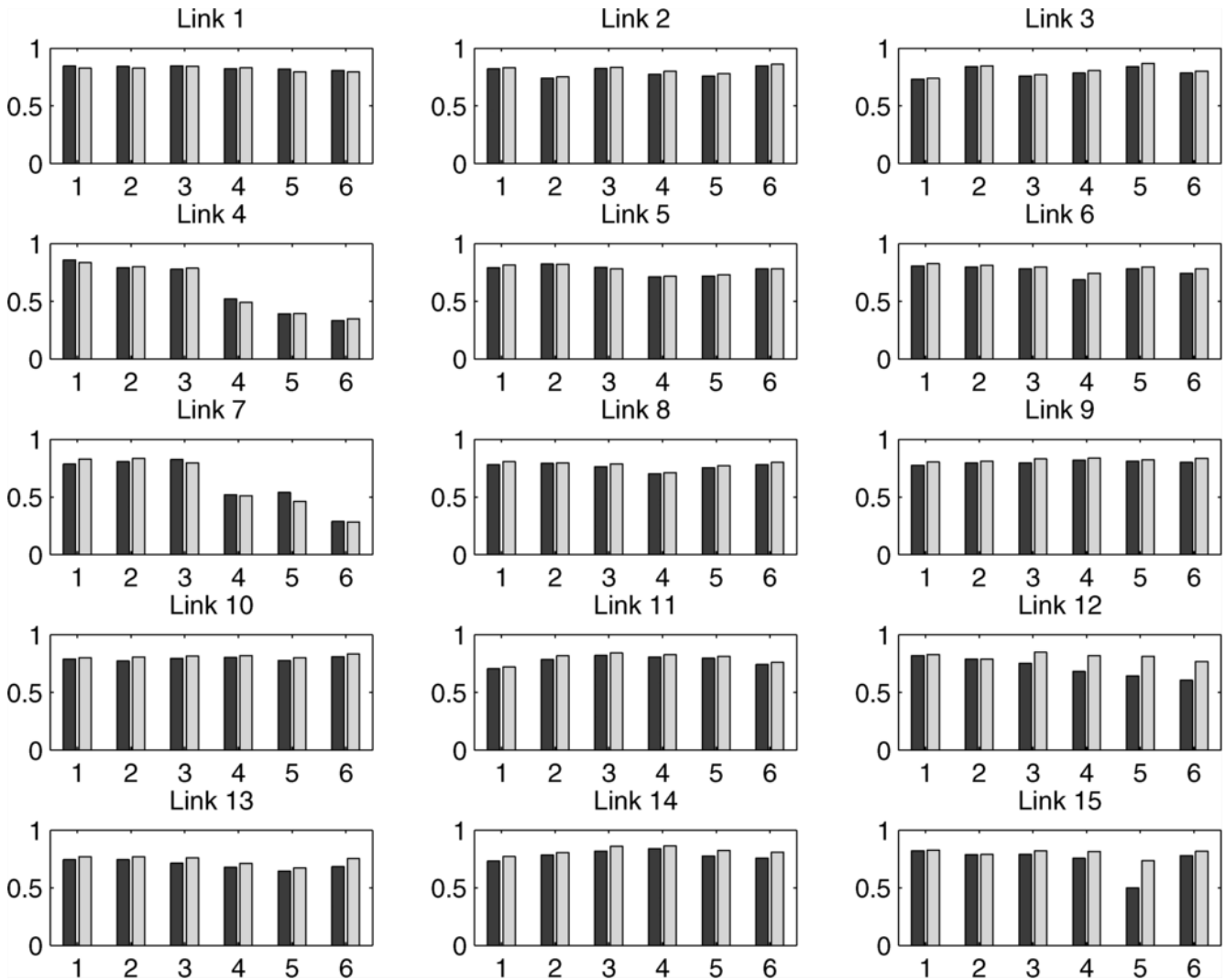

FIG. 2. *Probability of one unit or less delay on each link at each of six time periods. Dark is the estimate and light is the observed true value.*

The bin size for discretizing the link delays was chosen to be $q = 0.005$ second, about the 0.8 quantile on most links during the initial (stable) period. Depending on the application of interest, this can be changed to take into account the appropriate time scale, that is, for VoIP, video-conferencing or gaming. The inverse problem of obtaining the link delay distributions from end-to-end path delay measurements was solved every ten minutes to mimic a monitoring scenario. At its longest, the estimation procedure took about 40 seconds on a G4 Powerbook.

Figure 2 shows the probability of a delay of one unit (0.005 s) or less with the actual internal link-level data in light (which we can observe in this case since it is a simulation) as well as the estimated link-level delays in dark using tomography techniques in [8]. As we can see, the estimated distributions do a very good job of tracking the actual delay at all links of the network. Also, the changes on links 4 and 7 are captured well, and there seems very little or no downstream effect on the estimation for the daughter links of 7.

Formal monitoring procedures based on EWMA procedures are discussed in [12] for the loss problem. Initial results are given for the delay problem in [8]. We are currently studying various monitoring procedures and their properties in more detail. Researchers at Avaya Labs have also developed and implemented some quick and easy methods for visually monitoring the networks.

## ACKNOWLEDGMENTS

This research was supported in part by NSF Grants IIS-99-88095, CCR-03-25571, DMS-02-04247 and DMS-05-05535.